\documentclass{article}
\usepackage{graphicx}
\usepackage{amsmath,amssymb,mathtools}
\usepackage{graphics,color,array,calc,rotating,epsfig,psfrag}
\numberwithin{equation}{section}
\usepackage{cite}
\usepackage{bm}
\usepackage{dcolumn}
\usepackage{float}
\usepackage{subfigure}
\usepackage{amsfonts}
\usepackage[mathscr]{eucal}
\def\be{\begin{equation}} \def\ee{\end{equation}}
\def\bea{\begin{eqnarray}} \def\eea{\end{eqnarray}}

\newcommand{\nn}{\nonumber}
\newcommand{\MeijerG}[7]{G \begin{smallmatrix} #1 & #2 \\ #3 & #4 \end{smallmatrix} \left( \begin{smallmatrix} #5 \\ #6 \end{smallmatrix} \middle\vert #7 \right) }
\begin{document}
\baselineskip 18pt%
\begin{titlepage}
\vspace*{1mm}%
\hfill%
\vspace*{15mm}%
\hfill
\vbox{
    \halign{#\hfil         \cr
          } 
      }  
\vspace*{20mm}

\begin{center}
{\large {\bf  Non-commutative effects on gravitational measurements}}\\
\vspace*{5mm}
{ Majid Karimabadi, S. A. Alavi \footnote{s.alavi@hsu.ac.ir}, Davood Mahdavian Yekta}\\
\vspace*{0.2cm}
{$^{}$ Department of Physics, Hakim Sabzevari University, P.O. Box 397, Sabzevar, Iran}\\
\vspace*{1cm}
\end{center}

\begin{abstract}
The general theory of relativity is currently the accepted theory of gravity and
as such, a large repository of test results has been obtained since its inception in 1915.
However, in this paper we only focus on what are considered as the main tests but in non-commutative (NC) geometries. Using the coordinate coherent state formalism, we calculate the gravitational redshift, deflection, and time delay of light, separately, for NC inspired {\em Schwarzschild} and {\em Riessner-Nordstr\"{o}m} black holes. We show the NC predictions have different behavior than the general relativity. We also provide an upper bound on the NC parameter by comparing the NC corrections with the accuracies of gravitational measurements in the case of typical primordial black holes produced in the early universe. In this regard, we use observational data for the scale factor $a$ at the end of inflation to obtain physical satisfactory results.
\end{abstract}

\end{titlepage}
%
\section{Introduction}

Einstein's general theory of relativity (GR) changed fundamentally our understanding of space-time, mass, energy, and gravity. It can be used to predict many new features from the existence of black holes, one of the most interesting and mysterious objects in the universe to the gravitational lensing, redshift, and time delay, i.e., the three profound implications of GR\cite{Weinberg}.
Gravitational redshift is a very useful tool in astrophysics. It helps us to test our knowledge of the structure of those stars whose internal structures are  different from the sun and other normal stars\cite{Hetherington,Holberg}.

Another important prediction of this theory is the deflection of light rays passing close to a massive body. This phenomenon is called gravitational lensing, confirmed by Eddington for the first time \cite{Dyson:1920cwa}. It is still one of the major tools of cosmology\cite{Peebles:2004qg}, astrophysics\cite{Fomalont:1976zz,Fomalont:2009zg} and astronomy \cite{Bull:2015lja}. It is also useful for probing extra solar planets and detecting the presence and distribution of dark matter which act as sources of the gravitational field \cite{Song:2008xd}. Statistical study of lensing data provides very useful insight into the structural evolution of galaxies \cite{Rusin:2002tq,Reyes:2010tr}.\footnote{So far GR has been tested on scales smaller than an individual galaxy, e.g., orbits of planets in our solar system and the motion of stars around the center of the Milky way, but recently, it is shown in Ref. \cite{NP} that  GR also works well on galactic scale. Using data from the Hubble space telescope it was found that a nearby galaxy dubbed ESO 325-G 004 is surrounded by a ring like structure known as Einstein ring, which is an evidence for gravitational lensing. This precise test of GR on a galactic scale excludes some of the alternative theories of gravitation.}

Moreover, gravitational time delay measures the amount of time elapsed between two events by observers at different distances from a gravitational mass. To sum up, the clock placed in a higher gravitational potential region will run slower. Gravitational time delay has been confirmed by the Pound–Rebka experiment \cite{Pound:1960zz}, observations of the spectra of the white dwarf Sirius B, and experiments with time signals sent to and from Viking 1 Mars lander. Time delay corrections are also very important in Global Positioning Systems (GPS) \cite{Ashby:2003vja}.

On the other hand, motivated by string theory \cite{Seiberg:1999vs,Ardalan:1998ce}, generalized uncertainty principle \cite{Kempf:1994su}, quantum gravity \cite{AmelinoCamelia:2008qg}, Lorentz violation \cite{Colladay:1998fq,Carroll:2001ws}, and etc., the idea of NC spacetime has drawn quite a lot of interest in a wide range of areas from condensed matter physics\cite{Santos:2018ler,Hosseinzadeh:2015jra} to high energy physics\cite{Chaichian:2000si}, cosmology and astrophysics \cite{Connes:1997cr,Nasseri:2004ia,Alavi:2004aq}. In a number of scenarios in string theory based on the large extra-dimensions there is a bound on the unification scale, thus, if NC geometry as a promising alternative as suggested by some theories including string theory, the corresponding length scale should be satisfied in this bound\cite{Antoniadis:1998ig,ArkaniHamed:1998rs}. These ideas offer exciting, near future, possibility of experimentally probing both NC and quantum gravity effects.

The NC space is realized by the coordinate operators satisfying:
\be\label{nc} [x_{i},x_{j}]=i \theta_{ij}\,,\ee
where $\theta_{ij}$ is a constant anti-symmetric matrix. The relations (\ref{nc}) are not invariant under naive Lorentz transformations either. But they are invariant under a deformed Lorentz Symmetry \cite{Chaichian:2004za}, in which the coproduct on the Lorentz group is deformed while the group structure is kept intact. The main mathematically correct, but physically hard to implement, formalism to study the non-commutativity is the star-product or Moyal product approach \cite{Seiberg:1999vs,Harikumar:2006xf}. But there is an alternative approach based on the coordinate coherent states \cite{Snyder:1946qz,Doplicher:1994tu}, in which the concept of point-like particle becomes physically meaningless and must be replaced by its best approximation, i.e., a minimal width Gaussian distribution of mass and charge. In fact, the characteristic length scale of this system is given by the matter distribution width $\sqrt{\theta}$.
There has also been a growing interest in possible gravitational observable consequences of non-commutativity of space coordinates, specially the behavior of black holes in NC spaces.  Solutions of Einstein equations with such smeared sources give new kind of regular black holes in four \cite{Nicolini:2005vd,Ansoldi:2006vg,DiGrezia:2006rw,Nozari:2008rc,Banerjee:2008du,Wei:2015dua,Dymnikova:2016nlb} and higher dimensions \cite{Rizzo:2006zb,Gingrich:2010ed,Nouicer:2007cw,Nouicer:2007pu,Ye:2017blp,Ghosh:2017odw}. (For a general review of this formalism see \cite{Nicolini:2008aj})

In this paper we are curious to provide an upper bound for NC parameter $\theta$.
It is worth mentioning that a lower bound for NC parameter $\theta$ has been obtained in some papers through the study of black holes thermodynamics in NC spaces as  $\sqrt{\theta}\sim 10^{-1}\,l_{P}\sim10^{-34} cm$, see e.g., \cite{Nicolini:2005vd,Alavi:2009tn,Kim:2010xt}. To determine the upper bound we use the three aforementioned predictions of GR for {\em Schwarzschild} (Sch) and {\em Reissner-Norstr\"{o}m} (RN) black holes inspired by NC geometry based on the coordinate coherent state formalism. In the other words, we approximate that the accuracy of our calculations for some typical primordial black holes produced in the early universe are in the order of the accuracy of gravitational measurements, then we determine a numerical value for the upper bound on $\theta$. It would be also of interest to display explicitly the deviation from GR in the case of measurements in NC space.

In addition, there is strong evidence from observational data that inflation\cite{Guth:1980zm,Linde:1981mu,Starobinsky:1982ee,Albrecht:1982wi}, a period of accelerated expansion in the very early history of the universe, happened when the energy scale of the universe was about three orders of magnitude lower than the planck scale.  Inflation stretches a region of Planck size into cosmological scales. So, at the end of inflation, physics at the Planck scale can leave its signature on cosmological scales. That is, if the spacetime is NC at very small length scales then such scales would have been stretched to possibly detectable cosmological sizes at the end of inflation\cite{Lizzi:2002ib,Brandenberger:2002nq,Huang:2003zp,Koivisto:2010fk,Soda:2012zm}. However, we should know the value of the cosmological scale factor $a$ when inflation ended. At the GUT energy scale ($10^{16} GeV$), the scale factor at the end of inflation is $a\sim 10^{-29}$. In order to get the physical distance as it would be measured by an observer at any time t, we should multiply the results by scale factor $a$.
The authors in Refs. \cite{Ade:2013zuv,K.:2014ssa} have obtained a similar constraint on the NC parameter using CMB data from PLANCK and also in \cite{Akofor:2008gv} using data from WMAP \cite{Komatsu:2008hk}, ACBAR\cite{Reichardt:2008ay} and CBI \cite{Mason:2002tm,Sievers:2005gj}.
With these ideas in mind, in this paper, we look for a constraint on the NC length scale from observational data.

\section{NC geometry in coordinate coherent state formalism}
In the original idea of Snyder in \cite{Snyder:1946qz} there is a minimal length scale $\sqrt{\theta}$ at  which the classical concept of smooth spacetime manifold breaks down. From this perspective the $\theta$ parameter is a length squared quantity defining
the scale where spacetime coordinates become non-commuting. In the coordinate coherent state formalism instead of using a point-like structure for the matter, a ``matter droplet'' description is usually used for the mass with a Gaussian distribution. Following the construction in \cite{Nicolini:2005vd,Ansoldi:2006vg} we define
\be\label{gd} \rho_{M}(r)=\frac{M}{(4\pi \theta)^{\frac32}} \exp{\left(-\frac{r^2}{4\theta}\right)}\,,\ee
\be\label{gdq} \rho_{Q}(r)=\frac{Q}{(4\pi \theta)^{\frac32}} \exp{\left(-\frac{r^2}{4\theta}\right)}\,,\ee
where $\theta$ is the NC parameter in (\ref{nc}). For an observer at large distance this smeared density looks like a small sphere of matter with radius about $\sqrt{\theta}$ of mass $M$ and charge $Q$, so the metric becomes Sch ($Q=0$) or RN black hole at large distances. In fact, they are solutions of Einstein-Maxwell equations in the presence of a static, spherically symmetric, minimal width, gaussian distribution
of mass and charge \cite{Ansoldi:2006vg}.
But in the intermediate region the metric is neither Sch nor RN and can be analytically written in terms of lower incomplete gamma or error functions as shown bellow. The value of the total mass and charge are given by the integration of this bell-shaped function over the whole space as
\be \label{mass} M_{\theta}=4\pi \int r^2 \rho_{M}(r) dr=M\,\textrm{erf}\!\left[\frac{r}{2\sqrt{\theta}}\right]-\frac{M r}{\sqrt{\pi\theta}} e^{-\frac{r^2}{4\theta}}\,,\ee
\be \label{charge} Q_{\theta}=Q \,\textrm{erf}\!\left[\frac{r}{2\sqrt{\theta}}\right]+\frac{Q r\, e^{-\frac{r^2}{4\theta}}}{\sqrt{\pi \theta}}\,,\ee
where ``erf'' is the error function with the following properties
\be \label{erf} \lim_{x\rightarrow \infty}\textrm{erf}(x)=1,\qquad \lim_{x\rightarrow \infty}\textrm{erfc}(x)=0\,. \ee

In this paper, we study stationary, spherically symmetric, asymptotically flat metrics whose line element is given by
\be\label{gm} ds^2=-B(r) dt^2+\frac{dr^2}{B(r)}+r^2 d\Omega_{2}^2\,,\ee
where for general Sch and  RN metrics we respectively have
\be\label{srn}B_{Sch}(r)=1-\frac{2 G M}{r}\,,\qquad B_{RN}(r)=1-\frac{2 G M}{r}+\frac{Q^2}{r^2}\,.\ee

If we insert the mass distribution (\ref{gd}) in the Einstein equations of motion, it can be seen that the solution is a Sch black hole \cite{Nicolini:2005vd} with
\be \label{ncss}  B_{Sch}^{NC}(r)=1-\frac{2 G M}{r}+\frac{2 G M}{r}\, \textrm{erfc}\!\left[\frac{r}{2\sqrt{\theta}}\right]+\frac{2 G M e^{-\frac{r^2}{4\theta}}}{\sqrt{\pi \theta}},\ee
and for both (\ref{mass}) and (\ref{charge}) in Einstein-Maxwell equations, we have RN metric \cite{Ansoldi:2006vg}
\bea\label{ncrn} B_{RN}^{NC}(r)&=&1-\frac{2 G M}{r}+\frac{2 G M}{r}\, \textrm{erfc}\!\left[\frac{r}{2\sqrt{\theta}}\right]+\frac{2 G M e^{-\frac{r^2}{4\theta}}}{\sqrt{\pi \theta}}
+\frac{Q^2\, \textrm{erfc}\!\left[\frac{r}{2\sqrt{\theta}}\right]^2}{r^2}\nn\\
&-&\!\!\frac{Q^2\,}{r \sqrt{2\pi \theta}}\left( \textrm{erfc}\!\left[\frac{r}{\sqrt{2\theta}}\right]- \textrm{erfc}\!\left[\frac{r}{2\sqrt{\theta}}\right]\right)-\frac{Q^2 e^{-\frac{r^2}{4\theta}}}{\sqrt{2}\pi\theta}\,.\eea
Since the non-commutativity parameter, if it is non-zero, should be very small compared to the length scales of the system, so the commutative limit is given by $\frac{r}{\sqrt{\theta}}\rightarrow \infty$ or equivalently $\frac{\sqrt{\theta}}{r}\rightarrow 0$.  It must be noted that in this limit we obtain the metrics described by (\ref{srn}).

\begin{figure}[H]
\centering
\includegraphics[width=7cm,height=4cm]{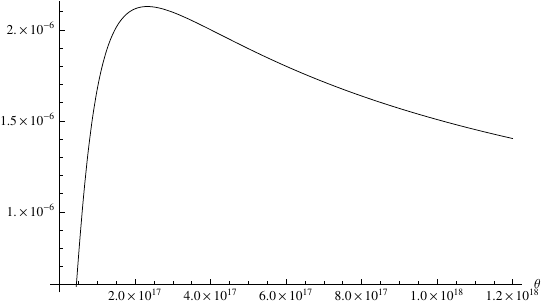}
\caption{{The last term of (\ref{ncss}).}}
\label{sh}
\end{figure}

As we know, for a typical gravitational system such as the sun, the value of the terms  including $\frac{2GM}{r}$ in (\ref{ncss}) are very small even for lowest value of $r$, which can be the Sch radius of the gravitational system, so in the next calculations we can ignore its higher orders without loss of generality. To prove that the last term in (\ref{ncss}) is very small, we have plotted this term in terms of different values of parameter $\theta$ for the Sch radius of the sun, $r_\circ$, as shown in Fig. (\ref{sh}). It is easy to check that for larger distances, i.e., $r> r_\circ$, the height of the peak decreases. Also, we can check this behavior for other gravitational systems, but we should pay attention that by increasing the mass, the radius of the Sch increases, too.

We have also applied similar consideration for the last two terms of relation
 (\ref{ncrn}) which are plotted in Fig.~(2a); $\frac{Q^2\,}{r \sqrt{2\pi \theta}}\left( \textrm{erfc}\!\left[\frac{r}{\sqrt{2\theta}}\right]- \textrm{erfc}\!\left[\frac{r}{2\sqrt{\theta}}\right]\right)$  and Fig.~(2b); $\frac{Q^2 e^{-\frac{r^2}{4\theta}}}{\sqrt{2}\pi\theta}$.
\begin{figure}[H]
\centering
\subfigure[]
{\label{rn1}\includegraphics[width=.4\textwidth]{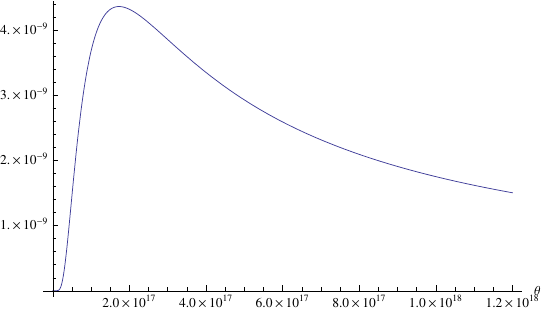}}
\subfigure[]
{\label{rn2}\includegraphics[width=.4\textwidth]{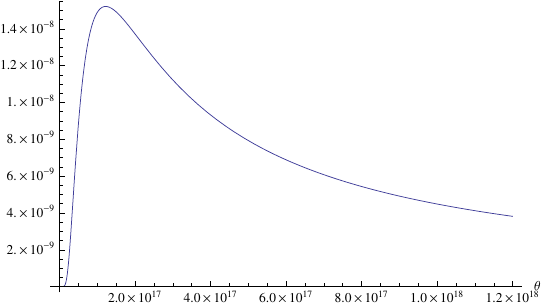}}
	    \caption{{The last two terms of (\ref{ncrn}).}}
\label{f1}
\end{figure}
In contradiction to the case of Sch solution in GR which has a single horizon, in NC space we have different possibilities \cite{Nicolini:2005vd}:
\begin{itemize}
\item For $\eta=\frac{M}{\sqrt{\theta}}<1.9$ there is no horizon for (\ref{ncss}) shown by red solid curve in Fig.~(\ref{mss})
\item For $\eta=\frac{M}{\sqrt{\theta}}=1.9$ there is a degenerate horizon in $x=\frac{r}{\sqrt{\theta}}=3$ shown by blue curve in Fig.~(\ref{mss}).
\item For $\eta=\frac{M}{\sqrt{\theta}}>1.9$ there are two distinct horizons shown by green curve in Fig.~(\ref{mss}).
\end{itemize}

\begin{figure}[H]
\centering
\includegraphics[width=8cm,height=5cm]{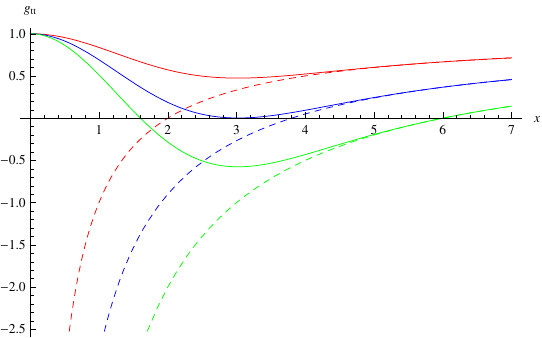}
\caption{{The $tt$ component of Sch metric in commutative (dashed) and non-commutative (solid) spaces for different values of $\frac{M}{\sqrt{\theta}}$.}}
\label{mss}
\end{figure}
\begin{figure}[H]
\centering
\includegraphics[width=6cm,height=4cm]{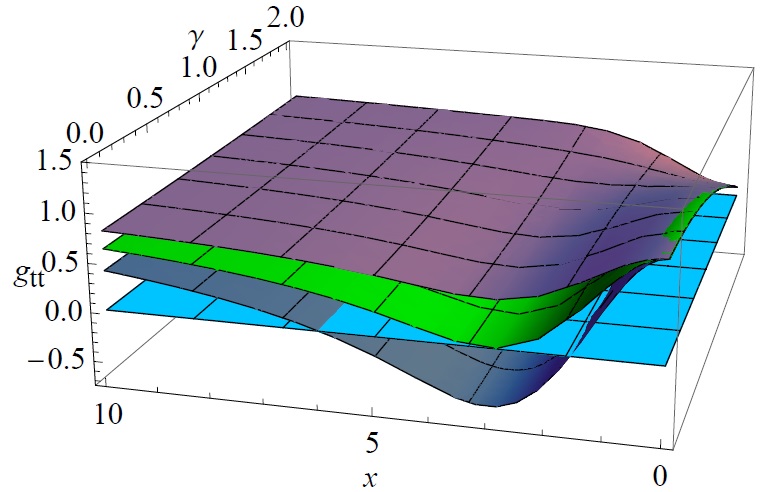}
\caption{{The Reissner-Nordstrom metric in NC space vs. $x=\frac{r}{\sqrt{\theta}}$ and $\gamma=\frac{Q}{\sqrt{\theta}}$ for different values of $\frac{M}{\sqrt{\theta}}$.}}
\label{ncrnm}
\end{figure}

We have also plotted the $tt$ component of charged solution metric (\ref{ncrn}) in Fig.~(\ref{ncrnm}). As seen in the limit $\frac{Q}{\sqrt{\theta}}\rightarrow 0$, we obtain the plots for Sch solution in NC space shown in Fig.~(\ref{mss}).

\subsection{Modified Newton's law}
In this subsection, we investigate modification of the Newton's law in coordinate coherent state formalism by using modified metric of Sch geometry. As done in \cite{Taylor}, one can first obtain the gravitational field as acceleration of a test particle around the massive object, such as a Sch black hole, and then multiply it by the mass of particle to reproduce the Newton's law of gravity \cite{Ali:2015zua}. In this paper, we use another approach in which we first calculate the energy of the particle and then use its derivative to give Newton's law. In general, the 4-velocity of a particle in curved space-time is given by $u^{\mu}=\frac{dx^{\mu}}{ds}$ such that $u^{\mu}u_{\mu}=-1$, where $s$ parameterizes the path of particle and is not necessarily identified with the time coordinate and that $ds^2=-g_{\mu\nu}dx^{\mu} dx^{\nu}$. Also the corresponding 4-momentum of particle of mass $m$ is $p^{\mu}=m u^{\mu}$. Then the conserved energy of particle is actually defined as $E=-p^{\mu}\xi_{\mu}$ where $\xi^{\mu}$ is a time-like killing vector field and the negative sign coming from the Lorentzian signature of the metric (\ref{gm}). So, by choosing $\xi^{\mu}=t^{\mu}=(1,0,0,0)$ the energy of particle in a curved space with non-rotating, spherical symmetry is
\be\label{mnl1} E=-m t_{\mu} u^{\mu}=-m g_{\mu\nu} t^{\nu} u^{\mu}=-m g_{00} t^{0} u^{0}=m B(r) \frac{dt}{ds}\,.\ee
 When the particle is at distance $r$, by substituting $dr=d\Omega=0$ in the metric (\ref{gm}) we have $ds^2=-B(r) dt^2$, thus the energy becomes
\be \label{mnl2} E=m\sqrt{-B(r)}\,. \ee
Now by differentiating with respect to $r$, we obtain the gravitational force on the test particle as
\be \label{mnl3} F=m\frac{B'(r)}{2\sqrt{-B(r)}}\,.\ee

Substituting $B(r)$ from the relations (\ref{ncss}) and (\ref{ncrn}) in (\ref{mnl3}), the gravitational force for NC inspired Sch and RN metrics are given by
\be\label{mnl4} F^{NC}_{Sch}=\frac{G m M}{r^2}\,\left(1-\frac{r^3 e^{-\frac{r^2}{4\theta}}}{2\sqrt{\pi} \theta^{3/2}}-\frac{r e^{-\frac{r^2}{4\theta}}}{\sqrt{\pi\theta}}-\text{erfc}\left[\frac{r}{2\sqrt{\theta}}\right]\right)\,,\ee
\bea\label{mnl5} F^{NC}_{RN}&\!\!\!=\!\!\!&F^{NC}_{Sch}-\frac{mQ^2}{r^3}\Big(1-\frac{r^4 e^{-\frac{r^2}{4\theta}}}{4\sqrt{2}\, \pi \theta^{2}}+\frac{r^2 e^{-\frac{r^2}{2\theta}}}{2\pi \theta}-\frac{r^2 e^{-\frac{r^2}{4\theta}}}{2\sqrt{2}\pi \theta}\nn\\
&\!\!\!-\!\!\!&\frac{r e^{-\frac{r}{2\theta}}}{\sqrt{\pi \theta}}\,\text{erfc}\left[\frac{r}{2\sqrt{\theta}}\right]+\frac{r }{2\sqrt{2\pi \theta}}\,\text{erfc}\left[\frac{r}{2\sqrt{\theta}}\right]
+\text{erfc}\left[\frac{r}{2\sqrt{\theta}}\right]^2-\frac{r }{2\sqrt{2\pi \theta}}\,\text{erfc}\left[\frac{r}{\sqrt{2\theta}}\right]\Big)\,,
\eea
where in the limit $\frac{\sqrt{\theta}}{r}\rightarrow 0$ they generally lead to Newton's law $F=\frac{GMm}{r^2}$. We have expanded the relativistic term $[-B(r)]^{-1/2}$ and only saved the leading term of order 1, since this term produces higher order corrections in $GM/r$ and $Q^2/r^2$. One can use the relations (\ref{mnl4}) and (\ref{mnl5}) to study the gravitational atoms formed by black holes which are candidates for Dark matter.
\section{Gravitational redshift}
When the light moves upwards against gravitational field, it loses some of its energy, so it  undergoes a redshift \cite{Weinberg}. Indeed, the light has a transition from high frequency (short wavelength) to a light with low frequency (long wavelength). So there is a shift in the spectral lines of light due to gravity given by
\be \label{red} z=\frac{\lambda_2-\lambda_1}{\lambda_1}=\frac{\omega_1-\omega_2}{\omega_2},\ee
which is known as gravitational redshift of light and is a direct consequence of Einstein General relativity theory. Now suppose that light was emitted from radius $r_1$ and received at $r_2$, so in the case of the general metric given in (\ref{gm}) the maximum value by which the light gets redshifted is \cite{Weinberg}
\be\label{reds2} z=\frac{\omega_1}{\omega_2}\Big|_{max}-1=\sqrt{\frac{B(r_2)}{B(r_1)}}-1\,,\ee
where $\omega_2$ and $\omega_1$ are the frequency received by the observer and the frequency emitted by the source, respectively. Using Eqs. (\ref{srn}) and (\ref{reds2}), we obtain the gravitational redshift for the Sch and RN black holes in GR, respectively,
\be\label{srnreds} z=\sqrt{\left(1-\frac{2GM}{r_2}\right) \left(1-\frac{2GM}{r_1}\right)^{-1}}-1\,,\quad z=\sqrt{\left(1-\frac{2GM}{r_2}+\frac{Q^2}{r_2^2}\right)\left(1-\frac{2GM}{r_1}+\frac{Q^2}{r_1^2}\right)^{-1}}-1\,.\ee

\subsection{Gravitational redshift in NC space}
\begin{figure}[H]
\centering
\includegraphics[width=8cm,height=5cm]{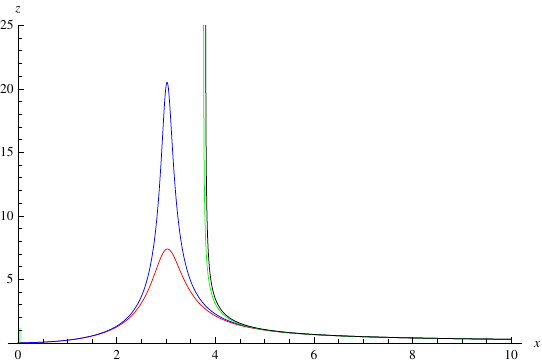}
\caption{{ The redshift of NC inspired Sch (blue curve) and RN (red curve)  metric vs. $x$ for $\frac{M}{\sqrt{\theta}}=1.9$ and $\frac{Q}{\sqrt{\theta}}=0.5$. The green and black curves are the Sch and RN metrics in GR, respectively. }}
\label{srnred}
\end{figure}

In this subsection, we first study the behavior of gravitational redshift factor for the non-commutative Sch and RN backgrounds, and then we find upper bounds for the NC parameter $\theta$ by comparing the results with the accuracy of the measurements. The redshift for a NC Sch black hole measured by an asymptotic observer, $r_2\rightarrow \infty$, is given by substituting the component (\ref{ncss}) in the relation (\ref{reds2})
\be\label{ncsred} z=\left(1\!-\!\frac{2 G M}{r_1}\!+\!\frac{2 G M \textrm{erfc}\!\left[\frac{r_1}{2\sqrt{\theta}}\right]}{r_1}+\frac{2 G M e^{-\frac{r_1^2}{4\theta}}}{\sqrt{\pi \theta}}\right)^{-\frac12} \!-\!1,\ee
and for a NC RN black hole by inserting (\ref{ncrn}) in the relation (\ref{reds2}) we have
\bea \label{ncrnred} z&\!\!\!=\!\!\!&\Big[1-\frac{2 G M}{r_1}+\frac{2 G M\textrm{erfc}\!\left[\frac{r_1}{2\sqrt{\theta}}\right]}{r_1} +\frac{2 G M e^{-\frac{r_1^2}{4\theta}}}{\sqrt{\pi \theta}}
+\frac{Q^2\, \textrm{erfc}\!\left[\frac{r_1}{2\sqrt{\theta}}\right]^2}{r_1^2}\nn\\
&\!\!-\!\!&\frac{Q^2\,}{r_1 \sqrt{2\pi \theta}}\left( \textrm{erfc}\!\left[\frac{r_1}{\sqrt{2\theta}}\right]- \textrm{erfc}\!\left[\frac{r_1}{2\sqrt{\theta}}\right]\right)-\frac{Q^2 e^{-\frac{r_1^2}{4\theta}}}{\sqrt{2}\pi\theta}\Big]^{-\frac12}-1\,,\eea
where again in the limit $\frac{\sqrt{\theta}}{r_1}\rightarrow 0$ they lead to the GR redshifts (\ref{srnreds}) for $r_2\rightarrow \infty$.
We have plotted the redshifts for both the Sch (blue curve) and RN (red curve) metrics in NC and GR background (black and green curves) in Fig.~(\ref{srnred}) in terms of the scaled radial coordinate $x=\frac{r_1}{\sqrt{\theta}}$. As expected for far regions from the gravitational system, all of them tend to zero and there is no shift in the light wavelength. We should note that in Fig.~(\ref{srnred}) and the other figures in this paper, only the regions outside the event horizon are physically acceptable and there is no compelling evidences to analyze the behavior below the event horizon, though the black holes are regular in NC space. It can be also seen that in contrast to GR predictions, there is a finite extremum for the red shift value and this extremum occurs at $x=3$ (of course for the RN not exactly at $x=3$), which is consistent with the extremal limit, $\eta=1.9$, illustrated in Fig.~(\ref{mss}). On the other hand, this extremum value occurs at the degenerate horizon. Comparing the plots of the Sch and RN in Fig.~(\ref{srnred}), we can infer that the value of the extremum has decreased for the RN background metric relative to Sch's one. The authors in \cite{Nicolini:2009gw} have obtained a similar expression for gravitational redshift in the case of NC inspired dirty black holes as well.

\section{Gravitational deflection of light}
One of the other interesting predictions and early verifications of GR is the deflection of light caused by gravity which is also known as gravitational lensing. When the light passes close to a massive object such as a supernova or a black hole, it is deflected from its straight path which is given by the following relation\cite{Weinberg}
\be\label{def} \Delta\phi=2 \int^{\infty}_{r_{\circ}} \frac{1}{r\sqrt{B(r)}} \,\left(\frac{r^2}{r_{\circ}^2}\frac{B(r_{\circ})}{B(r)}-1\right)^{-\frac12} dr-\pi\,,\ee
where $r_{\circ}$ is the closest distance to the massive object. Calculating the integration yields to the following expressions for the Sch and RN metrics, in commutative space:
\be\label{srndef} \Delta\phi_{Sch}=\frac{4GM}{r_{\circ}}\,,\qquad \Delta\phi_{RN}=\frac{4GM}{r_{\circ}}-\frac{3\pi Q^2}{4 r_{\circ}^2}\,.\ee
\subsection{Deflection of light in NC spaces}
This part of work is devoted to the study of deflection of light (light bending) in the NC spaces. By substituting the metric (\ref{ncss}) in the relation (\ref{def}) we have
\bea\label{defsh1} \Delta\phi&=&2\,\int_{r_{\circ}}^{\infty} \Big[\frac{1}{r \sqrt{\frac{r^2}{r_{\circ}^2}-1}}-\frac{G M}{\left(\frac{r^2}{r_{\circ}^2}-1\right){}^{3/2} r_{\circ}^2}+\frac{G M r}{\left(\frac{r^2}{r_{\circ}^2}-1\right){}^{3/2} r_{\circ}^3}+\frac{G M}{r^2 \sqrt{\frac{r^2}{r_{\circ}^2}-1}}\nn\\
&&\qquad+\frac{G M \text{erfc}\left(\frac{r}{2 \sqrt{\theta }}\right)}{\left(\frac{r^2}{r_{\circ}^2}-1\right){}^{3/2} r_{\circ}^2}-\frac{G M r \text{erfc}\left(\frac{r_{\circ}}{2 \sqrt{\theta }}\right)}{\left(\frac{r^2}{r_{\circ}^2}-1\right){}^{3/2} r_{\circ}^3}-\frac{G M \text{erfc}\left(\frac{r}{2 \sqrt{\theta }}\right)}{r^2 \sqrt{\frac{r^2}{r_{\circ}^2}-1}}+\frac{G M r e^{-\frac{r^2}{4 \theta }}}{\sqrt{\pi } \sqrt{\theta } \left(\frac{r^2}{r_{\circ}^2}-1\right){}^{3/2} r_{\circ}^2}\nn\\
&&\qquad-\frac{G M r e^{-\frac{r_{\circ}^2}{4 \theta }}}{\sqrt{\pi } \sqrt{\theta } \left(\frac{r^2}{r_{\circ}^2}-1\right){}^{3/2} r_{\circ}^2}-\frac{G M e^{-\frac{r^2}{4 \theta }}}{\sqrt{\pi } \sqrt{\theta } r \sqrt{\frac{r^2}{r_{\circ}^2}-1}}\Big] \, dr-\pi\,,\eea
After integrating, we obtain the deflection of light caused by the Sch metric as follows
\be\label{defsh2} \Delta\phi_{Sch}^{NC}=\frac{4 G M}{r_{\circ}}-\frac{4 G M e^{-\frac{r_{\circ}^2}{4 \theta }}}{r_{\circ}}-\frac{G M r_{\circ} e^{-\frac{r_{\circ}^2}{4 \theta }}}{\theta }\,. \ee

In the limit $\frac{\sqrt{\theta}}{r_{\circ}}\rightarrow 0$, or equivalently $\frac{r_{\circ}}{\sqrt{\theta}}\rightarrow\infty$, the exponential term goes more  rapidly to zero than the other terms, so this relation gives the GR prediction (\ref{srndef}). In the case of RN metric described by (\ref{ncrn}), the integral (\ref{def}) leads to

\bea \label{rndef1} \Delta\phi&\!\!\!\!\!=\!\!\!\!\!&-\pi+2\,\int_{r_{\circ}}^{\infty} dr\,\Big[\frac{1}{r \sqrt{\frac{r^2}{r_{\circ}^2}-1}}-\frac{G M}{\left(\frac{r^2}{r_{\circ}^2}-1\right){}^{3/2} r_{\circ}^2}+\frac{G M r}{\left(\frac{r^2}{r_{\circ}^2}-1\right){}^{3/2} r_{\circ}^3}+\frac{G M}{r^2 \sqrt{\frac{r^2}{r_{\circ}^2}-1}}-\frac{G M \text{erfc}\left(\frac{r}{2 \sqrt{\theta }}\right)}{\left(\frac{r^2}{r_{\circ}^2}-1\right){}^{3/2} r_{\circ}^2}\nn\\
&\!\!\!\!\!+\!\!\!\!\!&\frac{G M r \text{erfc}\left(\frac{r_{\circ}}{2 \sqrt{\theta }}\right)}{\left(\frac{r^2}{r_{\circ}^2}-1\right){}^{3/2} r_{\circ}^3}-\frac{G M \text{erfc}\left(\frac{r}{2 \sqrt{\theta }}\right)}{r^2 \sqrt{\frac{r^2}{r_{\circ}^2}-1}}+\frac{G M r e^{-\frac{r^2}{4 \theta }}}{\sqrt{\pi } \sqrt{\theta } \left(\frac{r^2}{r_{\circ}^2}-1\right){}^{3/2} r_{\circ}^2}-\frac{G M r e^{-\frac{r_{\circ}^2}{4 \theta }}}{\sqrt{\pi } \sqrt{\theta } \left(\frac{r^2}{r_{\circ}^2}-1\right){}^{3/2} r_{\circ}^2}\nn\\
&\!\!\!\!\!-\!\!\!\!\!&\frac{G M e^{-\frac{r^2}{4 \theta }}}{\sqrt{\pi } \sqrt{\theta } r \sqrt{\frac{r^2}{r_{\circ}^2}-1}}\!+\!\frac{Q^2 \text{erf}\left(\frac{r}{2 \sqrt{\theta }}\right)}{2 \sqrt{2 \pi }  r_{\circ}^2 \sqrt{\theta } \left(\frac{r^2}{ r_{\circ}^2}-1\right)^{3/2}}\!-\!\frac{Q^2 \text{erf}\left(\frac{r}{\sqrt{2} \sqrt{\theta }}\right)}{2 \sqrt{2 \pi }  r_{\circ}^2 \sqrt{\theta } \left(\frac{r^2}{ r_{\circ}^2}-1\right)^{3/2}}\!-\!\frac{Q^2 r e^{-\frac{r^2}{4 \theta }}}{2 \sqrt{2} \pi   r_{\circ}^2 \theta  \left(\frac{r^2}{ r_{\circ}^2}-1\right)^{3/2}}\nn\\
&\!\!\!\!\!+\!\!\!\!\!&\frac{Q^2 r e^{-\frac{ r_{\circ}^2}{4 \theta }}}{2 \sqrt{2} \pi   r_{\circ}^2 \theta  \left(\frac{r^2}{ r_{\circ}^2}-1\right)^{3/2}}
\!+\!\frac{Q^2 r \text{erf}\left(\frac{ r_{\circ}}{\sqrt{2} \sqrt{\theta }}\right)}{2 \sqrt{2 \pi }  r_{\circ}^3 \sqrt{\theta } \left(\frac{r^2}{ r_{\circ}^2}-1\right)^{3/2}}\!-\!\frac{Q^2 r \text{erf}\left(\frac{ r_{\circ}}{2 \sqrt{\theta }}\right)}{2 \sqrt{2 \pi }  r_{\circ}^3 \sqrt{\theta } \left(\frac{r^2}{ r_{\circ}^2}-1\right)^{3/2}}\!+\!\frac{Q^2 \text{erfc}\left(\frac{r}{2 \sqrt{\theta }}\right)}{r^3 \sqrt{\frac{r^2}{ r_{\circ}^2}-1}}\nn\\
&\!\!\!\!\!-\!\!\!\!\!&\frac{Q^2 \text{erfc}\left(\frac{r}{2 \sqrt{\theta }}\right)}{ r_{\circ}^2 r \left(\frac{r^2}{ r_{\circ}^2}-1\right)^{3/2}}\!+\!\frac{Q^2 r \text{erfc}\left(\frac{ r_{\circ}}{2 \sqrt{\theta }}\right)}{ r_{\circ}^4 \left(\frac{r^2}{ r_{\circ}^2}-1\right)^{3/2}}\!+\!\frac{Q^2 \text{erfc}\left(\frac{r}{2 \sqrt{\theta }}\right)^2}{2  r_{\circ}^2 r \left(\frac{r^2}{ r_{\circ}^2}-1\right)^{3/2}}\!-\!\frac{Q^2 r \text{erfc}\left(\frac{ r_{\circ}}{2 \sqrt{\theta }}\right)^2}{2  r_{\circ}^4 \left(\frac{r^2}{ r_{\circ}^2}-1\right)^{3/2}}\!-\!\frac{Q^2 \text{erfc}\left(\frac{r}{2 \sqrt{\theta }}\right)^2}{2 r^3 \sqrt{\frac{r^2}{ r_{\circ}^2}-1}}\Big]. \eea
Integrating the first four lines gives the following expressions:
\bea\label{rndef2} \Delta\phi&\!\!\!\!=\!\!\!\!&\frac{4 G M}{r_{\circ}}-\frac{4 G M e^{-\frac{r_{\circ}^2}{4 \theta }}}{r_{\circ}}-\frac{G M r_{\circ} e^{-\frac{r_{\circ}^2}{4 \theta }}}{\theta }\nn\\
&\!\!\!\!-\!\!\!\!&\frac{r_{\circ} Q^2 e^{-\frac{r_{\circ}^2}{4 \theta }}}{2 \sqrt{2 \pi } \theta ^{3/2}}-\frac{Q^2 e^{-\frac{r_{\circ}^2}{2 \theta }}\Big(1- e^{-\frac{r_{\circ}^2}{4 \theta }}\Big)}{r_{\circ}\sqrt{2 \pi } \sqrt{\theta} }+\frac{Q^2\left(4\,\MeijerG{3}{0}{2}{3}{1,2}{0,\frac12,\frac32}{\frac{r_{\circ}^2}{4\theta}}-3\pi\right)}{4r_{\circ}^2}
\,,\eea
where the last term includes the $MeijerG$ function given in general textbooks
of Mathematics. But for the terms in the last line, i.e. the last five terms, we have no analytical integration and must perform them numerically. After numerical calculation, we see that the first two terms have opposite signs but with the same value for some typical values of $Q$, $M$, and $\theta$, so they cancel each other. This is also true for the second couple of terms, while in the case of the last term we observe that this is very small in comparison to the numerical values of other expressions in (\ref{rndef2}), at least at the order of $1\times 10^{-10}$, so we neglet this term. It should be noticed that there is a similar argument for different logical values of the parameters $Q$, $M$, and $\theta$.

\begin{figure}[H]
\centering
\includegraphics[width=7cm,height=4cm]{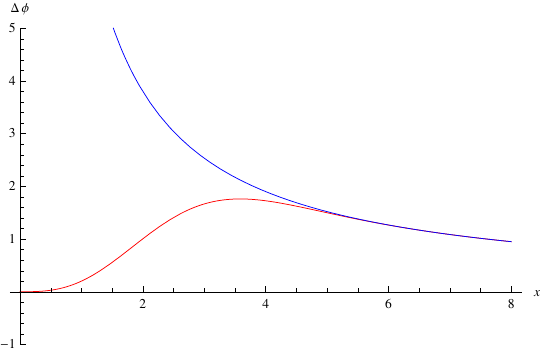}
\caption{{The Schwarzschild deflection for GR prediction (blue curve) and NC geometry (red curve). }}
\label{shd}
\end{figure}

Again, we can see that in the limit $\frac{\sqrt{\theta}}{r_{\circ}}\rightarrow 0$ the relation (\ref{rndef2}) goes to (\ref{srndef}),  i.e. the RN background in commutative space.
Now, in order to show the differences between commutative and non-commutative geometries, we plot the deflection relations. For example, in Fig.~(\ref{shd}) we have plotted the relations (\ref{srndef}) and (\ref{defsh2}) for the Sch black hole. The figure has been plotted for $\eta=\frac{M}{\sqrt{\theta}}=1.9$ and the maximum of the red curve is located around $x=\frac{r_{\circ}}{\sqrt{\theta}}=3$, which is the location of the degenerate horizon as mentioned in previous discussions. As it is observed, unlike commutative space, there is a finite deflection for light in NC geometry at the horizon and lower distances.

\begin{figure}[H]
\centering
\includegraphics[width=7cm,height=4cm]{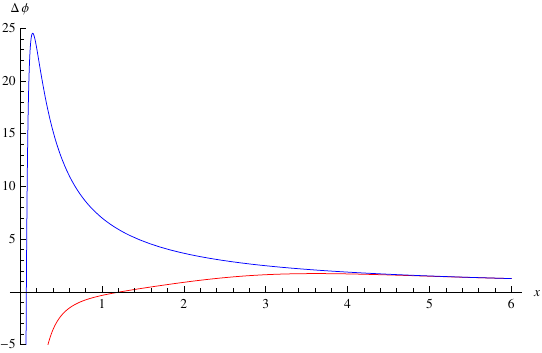}
\caption{{The RN deflection for GR prediction (blue curve) and NC geometry (red curve). }}
\label{rnd}
\end{figure}

We have also plotted the deflection of light which occurs in the vicinity of an RN metric with charge $\frac{Q}{\sqrt{\theta}}=0.5$ in Fig.~(\ref{rnd}). The blue curve is for GR given by (\ref{srndef}) and the red curve shows the results of RN metric in NC spaces as given by (\ref{rndef2}). Again it is seen that, there is a finite deflection for the light around the event horizon. It is worth mentioning that the results of both Sch and RN metrics coincide to the commutative case in the limit $\frac{\sqrt{\theta}}{r_{\circ}}\rightarrow 0$.

\section{Gravitational time delay}
Gravitational time delay is an actual difference of elapsed time between two events as measured by observers differently situated from a massive gravitational object. Since light has an effective mass, it is affected by gravity, so it follows a curved trajectory or a curved path when exposed to gravitational force. However, this is not actually what happens because gravity curves spacetime and light bends as a result of having to travel through that curved space-time. Since intense gravitational field attracts objects more towards itself (lensing), light bends more in a region of strong gravity. We also know that a straight path is always shorter than a curved path in terms of distance traversed. So, a light ray that bends more has to travel a longer distance to reach the destination and as a result it takes more time than a light ray that bends lesser. Einstein originally predicted this effect in his theory of relativity and it has since been confirmed by tests of GR. This phenomena was discovered and observed for the first time in 1964 by Shapiro \cite{Shapiro:1964uw}, which is also called Shapiro time delay.

Suppose that the light travels from $r = r_{1} $ to $r = r_{2}$, then the time delay is given by
\be\label{td}\Delta t_{max}=2\left(t(r_{\circ},r_{1})+t(r_{\circ},r_{2})-\sqrt{r_{\circ}-r_{1}}-\sqrt{r_{\circ}-r_{2}}\,\right)\,,\ee
where $ t(r_{\circ},r)$ is
\be\label{td1}  t(r_{\circ},r)= \int^{r}_{r_{\circ}} \frac{1}{{B(r)}} \,\left(1-\frac{r_{\circ}^2}{r^2}\frac{B(r)}{B(r_{\circ})}\right)^{-\frac12} dr\,,\ee
and $r_{\circ}$ is the distance of the closest approach to the typical gravitational system (a  massive object). Using Eqs. (\ref {srn}), (\ref {td}), and (\ref{td1}) the time delation for the light traveling from the source to the target in commutative geometry for Sch and RN black holes are
\be\label{td2}\Delta t_{max}^{Sch}=4MG(1+\ln{\frac{4r_{1}r_{2}}{r_{\circ}^{2}}})\,,\qquad \Delta t_{max}^{RN}=4MG(1+\ln{\frac{4r_{1}r_{2}}{r_{\circ}^{2}}})-3\frac{Q^{2}\pi}{r_{\circ}}\,,\ee
where $r_{\circ}$ is the event horizon of the massive object and $r_{1}$, $r_{2}$ are the distances of the source and the target from it, respectively.

\subsection{Shapiro time delay in NC space }
In order to study the Shapiro time delay in NC geometries, we will evaluate the integral (\ref{td1}) for the Sch (\ref{ncss}) and RN (\ref{ncrn}) metrics, separately. After integrating, we arrive at the following expression for Sch background
\bea\label{td3}t(r_{\circ},r)&\!\!\!\!=\!\!\!\!&2 G M \ln \left(\frac{\sqrt{r^2-r_{\circ}^2}+r}{r_{\circ}}\right)+\sqrt{r^2-r_{\circ}^2}+G M \sqrt{\frac{r_{\circ}-r}{r_{\circ}+r}}-\frac{r_{\circ}^2 G M e^{-\frac{r_{\circ}^2}{4 \theta }} \text{erf}\left(\frac{\sqrt{r^2-r_{\circ}^2}}{2 \sqrt{\theta }}\right)}{2 \theta }\nn\\
&\!\!\!\!-\!\!\!\!&\frac{r_{\circ} G M \text{erf}\left(\frac{r_{\circ}}{2 \sqrt{\theta }}\right)}{r \sqrt{1-\frac{r_{\circ}^2}{r^2}}}+\frac{G M \text{erf}\left(\frac{r}{2 \sqrt{\theta }}\right)}{\sqrt{1-\frac{r_{\circ}^2}{r^2}}}-3 G M e^{-\frac{r_{\circ}^2}{4 \theta }} \text{erf}\left(\frac{\sqrt{r^2-r_{\circ}^2}}{2 \sqrt{\theta }}\right)+\frac{r_{\circ}^2 G M e^{-\frac{r_{\circ}^2}{4 \theta }}}{\sqrt{\pi } \sqrt{\theta } r \sqrt{1-\frac{r_{\circ}^2}{r^2}}}\nn\\
&\!\!\!\!-\!\!\!\!&\frac{r_{\circ}^2 G M e^{-\frac{r^2}{4 \theta }}}{\sqrt{\pi } \sqrt{\theta } r \sqrt{1-\frac{r_{\circ}^2}{r^2}}}-\int^{r}_{r_{\circ}}\frac{2 G M \text{erfc}\left(\frac{r}{2 \sqrt{\theta }}\right)}{r \sqrt{1-\frac{r_{\circ}^2}{r^2}}} \, dr\,,\eea
where the first three terms come from GR and the other terms are corrections due to non-commutativity of space, so from (\ref{td}) the time delation is given by
\bea\label {td4} \Delta t_{max}^{NC-Sch}&\!\!\!\!=\!\!\!\!& \Delta t_{max}^{Sch}-\frac{r_{\circ}^2 G M e^{-\frac{r_{\circ}^2}{4 \theta }} \text{erf}\left(\frac{\sqrt{r_1^2-r_{\circ}^2}}{2 \sqrt{\theta }}\right)}{\theta }-\frac{r_{\circ}^2 G M e^{-\frac{r_{\circ}^2}{4 \theta }} \text{erf}\left(\frac{\sqrt{r_2^2-r_{\circ}^2}}{2 \sqrt{\theta }}\right)}{\theta }\nn\\
&\!\!\!\!-\!\!\!\!&6 G M e^{-\frac{r_{\circ}^2}{4 \theta }} \text{erf}\left(\frac{\sqrt{r_1^2-r_{\circ}^2}}{2 \sqrt{\theta }}\right)-6 G M e^{-\frac{r_{\circ}^2}{4 \theta }} \text{erf}\left(\frac{\sqrt{r_2^2-r_{\circ}^2}}{2 \sqrt{\theta }}\right)\nn\\
&\!\!\!\!-\!\!\!\!&\frac{2 r_{\circ} G M \text{erfc}\left(\frac{r_{\circ}}{2 \sqrt{\theta }}\right)}{r_1 \sqrt{1-\frac{r_{\circ}^2}{r_1^2}}}-\frac{2 r_{\circ} G M \text{erfc}\left(\frac{r_{\circ}}{2 \sqrt{\theta }}\right)}{r_2 \sqrt{1-\frac{r_{\circ}^2}{r_2^2}}}+\frac{2 G M \text{erfc}\left(\frac{r_1}{2 \sqrt{\theta }}\right)}{\sqrt{1-\frac{r_{\circ}^2}{r_1^2}}}+\frac{2 G M \text{erfc}\left(\frac{r_2}{2 \sqrt{\theta }}\right)}{\sqrt{1-\frac{r_{\circ}^2}{r_2^2}}}\nn\\
&\!\!\!\!+\!\!\!\!&\frac{2 r_{\circ}^2 G M }{\sqrt{\pi } \sqrt{\theta } r_1 \sqrt{1-\frac{r_{\circ}^2}{r_1^2}}}\left(e^{-\frac{r_{\circ}^2}{4 \theta }}-e^{-\frac{r_1^2}{4 \theta }}\right)+\frac{2 r_{\circ}^2 G M }{\sqrt{\pi } \sqrt{\theta } r_2 \sqrt{1-\frac{r_{\circ}^2}{r_2^2}}}\left(e^{-\frac{r_{\circ}^2}{4 \theta }}-e^{-\frac{r_2^2}{4 \theta }}\right)\nn\\
&\!\!\!\!-\!\!\!\!&4 \int^{r_1}_{r_{\circ}}\frac{2 G M \text{erfc}\left(\frac{r}{2 \sqrt{\theta }}\right)}{r \sqrt{1-\frac{r_{\circ}^2}{r^2}}} \, dr-4 \int^{r_2}_{r_{\circ}}\frac{2 G M \text{erfc}\left(\frac{r}{2 \sqrt{\theta }}\right)}{r \sqrt{1-\frac{r_{\circ}^2}{r^2}}} \, dr\,,\eea
where the first term, given in (\ref{td2}), shows the GR contribution and is equal to $240 \,(\mu s)$ or $72000\,(m)$ for a solar system in which $M$, $r_{\circ}$, $r_1$, and $r_2$ are the mass and radius of the sun, distances of earth and mercury from the sun, respectively. Since $r_1, r_2>r_{\circ}>>\sqrt{\theta } $, so according to (\ref{erf}) the contributions of all the terms in the third line goes to zero and also in the fourth line, the parentheses tend to zero because the terms inside have almost the same values but with opposite sign. Using the same argument one can arrive to the conclusion that the contributions of the two last terms are negligibly small.
Therefore the expression reduces to
\be\label {td8}
 \Delta t_{max}^{NC-Sch}= \Delta t_{max}^{Sch}-\left(\frac{r_{\circ}^2 G M e^{-\frac{r_{\circ}^2}{4 \theta }} }{\theta }+6 G M e^{-\frac{r_{\circ}^2}{4 \theta }}\right) \,\left[\text{erf}\left(\frac{r_1}{2 \sqrt{\theta }}\right)+\text{erf}\left(\frac{r_2}{2 \sqrt{\theta }}\right)\right].
\ee
Again employing (\ref{erf}) the final expression is as follows
\be\label {td9}
 \Delta t_{max}^{NC-Sch}= \Delta t_{max}^{Sch}-2\frac{r_{\circ}^2 G M e^{-\frac{r_{\circ}^2}{4 \theta }} }{\theta }-12 G M e^{-\frac{r_{\circ}^2}{4 \theta }}.
\ee

For the RN background, the calculation method is the same as the previous one. So, we refrain from providing details and only provide the final expression
\be\label {tdr1}\Delta t_{max}^{NC-RN}=\Delta t_{max}^{RN}-12 G M\,e^{\frac{r_{\circ}^2}{4\theta}}- \frac{2\,r_{\circ}^2 G M e^{-\frac{r_{\circ}^2}{4 \theta }}}{\theta }-\frac{2\,r_{\circ}^2 Q^2 e^{-\frac{r_{\circ}^2}{4 \theta }}}{\sqrt{2\pi}\theta^{3/2}}+\frac{4\,Q^2 e^{-\frac{r_{\circ}^2}{2 \theta }}}{\sqrt{2\pi \theta}}\,.\ee
One can check that the relations (\ref{td9}) and (\ref{tdr1}) tend to GR predictions in the limit $\frac{r_{\circ}}{\sqrt{\theta}}\rightarrow \infty$. In Figs.~(8a\,,\,8b), We have plotted both GR and NC predictions of dimensionless parameter $\xi=\frac{\Delta t_{max}}{r_{\circ}}$ instead of time delay for both the Sch and RN metrics. As seen in both figures, the predictions of GR and NC for large distances become consistent while in the vicinity of degenerate horizon at $x=\frac{r_{\circ}}{\sqrt{\theta}}=3$ the difference is obvious.
\begin{figure}[H]
\centering
\subfigure[{\em Schwarzschild metric}]
{\label{std1}\includegraphics[width=.4\textwidth]{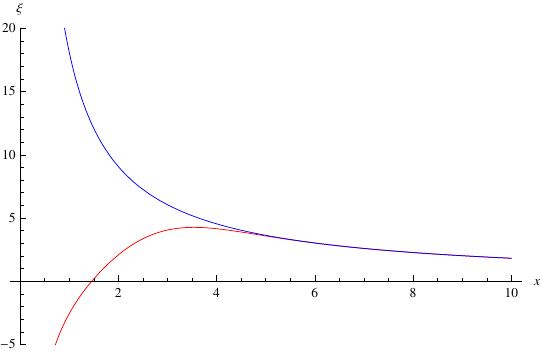}}
\subfigure[{\em Riessner-Nordstr$\ddot{o}$m metric}]
{\label{rntd2}\includegraphics[width=.4\textwidth]{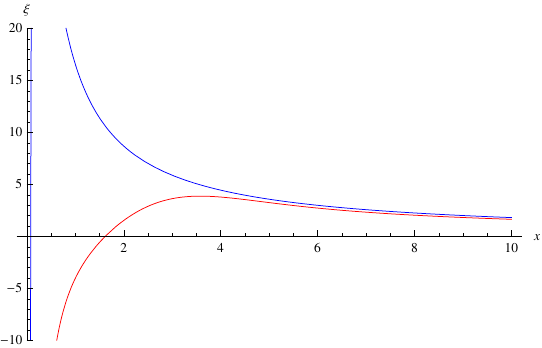}}
	    \caption{{Time delay for GR (blue curves) and NC (red curves) for $\eta = 1.9$ and $\frac{Q}{\sqrt{\theta}} =0.5$.}}
\label{tdf}
\end{figure}
\section{Upper bounds on $\theta$ from observational data}
According to the Hawking-Carr proposal \cite{Carr:1974nx}, the primordial black holes created in the early universe could have initial masses ranging from $10^{-8} kg$ to more than thousands of solar masses. However, primordial black holes with a mass lower than $10^{12} kg$ would have evaporated due to Hawking radiation in a time much shorter than the age of the universe, $8.33\times 10^{19}\,s$, so they cannot have survived until present time. But primordial black holes of a mass about $10^{12} kg$ were created in the Big Bang would be exploding today, so we should be able to observe some of them and this is why NASA's Fermi Gamma-ray Space Telescope satellite was designed. On the other hand, since the universe contains the cosmic microwave background radiation, in order for the black hole to dissipate, it must have a temperature higher than that of the present-day black body radiation of the universe, $2.7 K$. This implies that the mass of these black holes must be approximately near to the mass of the moon\cite{Kapusta:1999yy}.

This characteristic property of primordial black holes can help us to compare the corrections due to non-commutativity of space on different gravitational measurements, discussed in this paper, with their accuracies of the measurements, so that we can impose some bounds on the NC parameter $\theta$. The other most important thing that we should consider in our calculation is the accelerated expansion of the universe during inflation in which very small distances got stretched to cosmological sizes. That is, the non-commutativity of spacetime at very small length scales would have been stretched to possibly detectable cosmological sizes at the end of inflation. For cosmological application a natural choice of coordinates is the one where the time coordinate measures the proper time for an observer in a galaxy and the spatial coordinates measure the physical distances in such spatial slices. So what enters on the right-hand side of (\ref{nc}) is the physical $\theta$ which is constant throughout the history of the Universe. We carry out the computations here in the comoving frame and only in the end translate the result into the physical frame\cite{K.:2014ssa}. So to get the physical distance as it would be measured by an observer at any time t, we should multiply the results by cosmological scale factor $a\sim 10^{-29}$.

\begin{figure}[H]
\centering
\includegraphics[width=7cm,height=5cm]{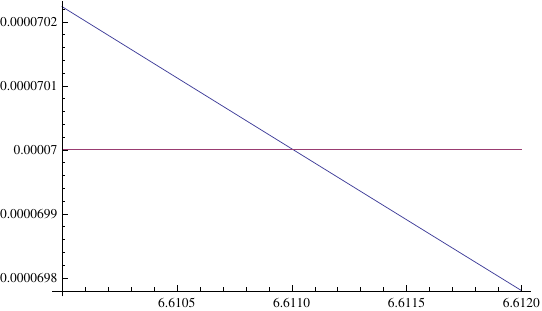}
\caption{Numerical comparison for redshift (\ref{bd1}), the horizontal axis is $ \frac{r_1}{\sqrt{\theta}}$.}
\label{rsb}
\end{figure}

For gravitational red shift, the most accurate measurement comes from Gravity Probe A \cite{Vessot:1980zz}. This accuracy was obtained by a satellite at an altitude of $10^7m$. By expanding (\ref{reds2}), we can separate GR term from NC corrections as follows
\be \label{fred} z_{NC}=z_{GR}\,\left[1+\left|\frac{z_{GR}+1}{z_{GR}}\right| \left(\frac{ G M \textrm{erfc}\!\left[\frac{r_1}{2\sqrt{\theta}}\right]}{(1-\frac{2GM}{r_1})\,r_1}+\frac{ G M e^{-\frac{r_1^2}{4\theta}}}{(1-\frac{2GM}{r_1})\sqrt{\pi \theta}}\right)\right]. \ee
The NC correction term should be smaller than the accuracy of the measurements which is  $7.0\times 10^{-5}$, so we have
\be\label{bd1} \left|\frac{z_{GR}+1}{z_{GR}}\right| \left(\frac{ G M \textrm{erfc}\!\left[\frac{r_1}{2\sqrt{\theta}}\right]}{(1-\frac{2GM}{r_1})\,r_1}+\frac{ G M e^{-\frac{r_1^2}{4\theta}}}{(1-\frac{2GM}{r_1})\sqrt{\pi \theta}}\right) \le 7.0\times 10^{-5}\,, \ee
where $z_{GR}$ is given in Eq. (\ref{srnreds}). Due to the complexity of the error function and the presence of an exponential term, solving this inequality analytically is a very difficult task or may be almost impossible, but instead we plot both sides and then find the value of $\theta$ in the intersecting point as depicted in Fig.~(\ref{rsb}). In plotting the figure, we have used the mass and radius of a typical micro black hole, $M G\sim 5\times 10^{-4} (m)$ and $r_1\sim1.5\times 10^{-3} (m)$. Therefore, according to the plot we obtain the upper bound as
\be \label{bd2} \alpha\ge 6.6 \Rightarrow \frac{1.5\times 10^{-3}}{\sqrt{\theta}}\ge 6.6 \Rightarrow \sqrt{\theta} \le 2\times 10^{-4} \,(m). \ee

The accuracy $7\times 10^{-5}$ is for sun's red shift measurement, but we can convince ourselves using this value, because the upper bound is not highly sensitive to the accuracy of the measurements which can be easily checked. This is in fact the case for other upper bounds which will be obtained later.

In the case of deflection of light, a number of measurements have been made in succeeding years, but there was no significant improvement until the advent of very-long-baseline interferometry (VLBI) \cite{Fomalont:2009zg}. VLBI is based on the interferometer implication in which there are different radio telescopes situated at distant locations and are utilized simultaneously. This technique creates a virtual radio telescope with a far larger baseline. VLBI uses a geometric method to measure the arrival time of radio waves from an astronomical source between different telescopes. For deflection of light, this project achieved an accuracy of $3\times 10^{-4}$. Thus, according to the NC definition in (\ref{defsh2}), we should have
\be\frac{|\delta\Delta\phi|}{\Delta\phi_{GR}}\le 3\times 10^{-4},\ee
where $\delta \Delta \phi=\Delta \phi_{NC}-\Delta\phi_{GR}$. Now, using  $r_{\circ}\sim 1.5\times 10^{-3} (m)$ and $G M\sim 5\times 10^{-4} (m)$ for a micro black hole, we have
\be\label{dlb} \frac{r_{\circ}}{4GM}(\frac{4 G M e^{-\frac{r_{\circ}^2}{4 \theta }}}{r_{\circ}}+\frac{G M r_{\circ} e^{-\frac{r_{\circ}^2}{4 \theta }}}{\theta })\le 3\times 10^{-4} \Rightarrow \sqrt{\theta} \le 2.3\times 10^{-4}\,(m). \ee

And finally for the time delay, the best measurement with high accuracy was obtained by the Cassini spacecraft \cite{Bertotti:2003rm} for time traveling of radio waves from earth to the spacecraft. This accuracy has the absolute value $2.3\times 10^{-5}$, so from the GR time delay in (\ref{td9}) we should have
\be \label{tdc} \left|\frac{\delta t_{max}^{NC-Sch}}{\Delta t_{max}^{Sch}}\right|\le 2.3\times 10^{-5} \Rightarrow \sqrt{\theta} \le 2.2\times 10^{-4}\,(m)\,, \ee
where $\delta t_{max}^{NC-Sch}=-\frac{2\, r_{\circ}^2 G M e^{-\frac{r_{\circ}^2}{4 \theta }} }{\theta }-12 G M e^{-\frac{r_{\circ}^2}{4 \theta }}$. It has been shown in \cite{Gruppuso:2005yw} that we can obtain a similar bound by computing the modification of the Newtonian potential due to the change of the Green’s function caused by the non-commutativity of the spacetime coordinates. Due to the verification of Newton's law up to a distance of the order of $200 \mu m$ \cite{Hoyle:2000cv}, the possible bound for NC parameter is $\theta <10^{-8} m^2$ which is of the order of bounds in relations (\ref{bd2}), (\ref{dlb}), and (\ref{tdc}).

The physical constraints on the NC parameter according to the above results are calculated by multiplying the value of $\theta$ by the scale factor $a\sim 10^{-29}$. So, we summarize our results in Tab. (\ref{tab1}). The order of these values are in accordance with the values in \cite{K.:2014ssa} for the NC parameter which obtained at the $2\!-\!\sigma$ confidence level.
\begin{table}[H]
\center
\begin{tabular}{|c|c|}\hline
     experiment  & physical bound($m$)\\ \hline
redshift& $\sqrt{\theta} \le 7.1\times 10^{-19}$ \\ \hline
 deflection& $\sqrt{\theta} \le 7.3\times 10^{-19}$\\ \hline
time delay& $\sqrt{\theta} \le 7.1\times 10^{-19}$ \\
 \hline
\end{tabular}
\caption[]{{Upper bounds on $\theta$ for different tests of GR}}
\label{tab1}
\end{table}


\section{Conclusion}

In this paper, we studied the three most important predictions of Einstein’s general relativity in NC spaces. Using some Gaussian distributions of the mass and charge, as physical parameters of these black holes in NC space, the metrics of Sch and RN are studied in coordinate coherent state formalism. In contrast to the commutative metrics which have an event horizon for each none-zero value of mass, we showed that in NC spaces, the existence of horizon tightly depends on the ratios $\frac{M}{\sqrt{\theta}}$ and $\frac{Q}{\sqrt{\theta}}$. Though we have only an event horizon for a commutative space, in NC case it is possible to have two different horizons at which $B(r)$ vanishes.
We have also obtained exact expressions for the gravitational redshift in both geometries as a function of $r$ and $\theta$ represented by (\ref{ncsred}) and (\ref{ncrnred}). As an important result which has been depicted in Fig.~(\ref{srnred}), in spite of GR in which the redshift factor does not have an extremum, in NC case we arrived at a finite extremum value in which light might shift to the red wavelength. It could even be seen from Fig.~(\ref{srnred}) that for the charged solution this value suppressed near the local degenerate horizon.

In addition, we have calculated the value of deflection of light (gravitational lensing) in the framework of NC spaces, which is different from GR prediction, that is, there is a maximum value for deflection of light near the horizon in NC solutions shown in Figs.~(\ref{shd}) and (\ref{rnd}). The third issue discussed in this paper is the gravitational time delay in NC space. As is obvious from the figures, in all of the predictions, the results of NC coincide with the commutative cases for large values of $x=\frac{r}{\sqrt{\theta}}$ while the differences appear in the vicinity of the degenerate horizon at $x=3$.
Another interesting result is that we have found upper bounds for NC parameter $\theta$  using the accuracies of gravitational measurements and observational data for cosmological scale factor. The physical results were obtained for some typical primordial black holes whose masses are near the mass of the moon and are given in Tab.(\ref{tab1}). An additional notable feature of these calculations is that the range of the NC parameter given in Tab.(\ref{tab1}) is also consistent with the lower bound for energy scale of $TeV$ that was obtained in a number of works in Refs. \cite{Horvat:2010sr,Horvat:2010km,Bilmis:2012sq} (see also Refs. in). These calculations can be generalized to higher dimensional theories of gravity in future studies \cite{New}. In general, the astrophysical black holes are of the Kerr type and one can apply similar calculations to this kind of backgrounds. There are rotating Kerr black hole solutions in NC geometries as in Refs. \cite{Modesto:2010rv,Singh:2017bwj}, but to the best of our knowledge the gravitational measurements have not been studied in NC Kerr metric. However, the calculation of the light deflection has been carried out in commutative (GR) Kerr black holes in Refs. \cite{Kraniotis:2005zm,Jusufi:2018jof}.
\section*{Acknowledgment}
We would like to thanks K. Nozari, A. Ghodsi, M. Ghominejad, M. Motahharfar, P. Mohseni, and A. Moulavi-Nafchi for careful reading the manuscript and for their valuable comments.

\end{document}